\documentclass{sig-alternate-10pt}
\usepackage[margin=1in]{geometry} 

\usepackage{graphicx}
\usepackage{latexsym}
\usepackage{amsmath}
\usepackage{float}
\usepackage{xspace}
\usepackage{algorithm}
\usepackage{algorithmic}
\usepackage{algorithmwh}
\usepackage{amssymb}
\usepackage{subfigure}
\usepackage{url}

\newdef{example}{Example}

\begin{document}
\pagestyle{empty}

%

\title{Privacy Preserving Web Query Log Publishing: A Survey on Anonymization Techniques}


\numberofauthors{1} 
%
\author{
%
%
\alignauthor
Amin Milani Fard\\
       \affaddr{Simon Fraser University, Burnaby, Canada}\\
       \affaddr{University of British Columbia, Vancouver, Canada}\\
       \email{milanifard@cs.sfu.ca}
}

\maketitle
\begin{abstract}
Releasing Web query logs which contain valuable information for research or marketing, can breach the privacy of search engine users. Therefore rendering query logs to limit linking a query to an individual while preserving the data usefulness for analysis, is an important research problem. This survey provides an overview and discussion on the recent studies on this direction.
\end{abstract}

\section{Introduction}
\label{one}

Web search queries are generally stored by search engines for the purpose of improving result ranking, query refinement, user modeling, language-based applications, and sharing data for academic research or commercial needs \cite{Cooper08}. On the other hand, releasing such data without proper anonymization can seriously breach the privacy of search engine users. In 2006 the America Online (AOL) query log data of 650k users over three months, was released after removing all explicit identifiers of searchers as shown in Figure 1 \cite{Barbaro06}. Shortly after that, the searcher No. 4417749 was traced back to the 62-year-old widow Thelma Arnold who lives in Lilburn. This scandal made data publishers reluctant in providing researchers with public anonymized query logs \cite{Hafner06}. Since then, an important research problem opened on rendering Web query log data to limit linking a query to a specific individual while the data is still useful for analysis. 

\begin{figure*}[t]
\small{
\begin{tabular}{|l|l|l|l|l|l|}
\hline
\textbf{AnonID} & \textbf{QueryContent} & \textbf{QueryTime} & \textbf{ItemRank} & \textbf{ClickURL}\\ \hline
303787 & hiv testing in hillsborough county ruskin florida & 2006-03-30 14:04:56 & 3 & http://www.ncsddc.org \\\hline
\end{tabular}
}
\caption{Sample query log from AOL released data [3]}
\end{figure*}

\subsection{Privacy-Preserving Data Publishing}

Researchers in the field of \textit{privacy-preserving data publishing} focus on designing techniques to publish data as useful as possible while preserving the privacy of individuals \cite{Fung10}. Publishing data instead of publishing data mining results is much more useful and interesting because many other analysis can be done on such data. Thus the published data should be potentially useful for many data analysis objectives which makes privacy-preserving data publishing challenging.

The process of \textit{anonymization} \cite{Cox80}\cite{Dal86} refers to hiding the identity (or sensitive information) of individuals. Removing explicit identifiers (such as name) is not effective since non-identifying personal data (such as age, gender, zipcode) can be combined with publicly available data to identify an individual \cite{Sweeney02}. The combination of such non-explicit identifiers are called the \textit{quasi-identifier} (QI) attributes \cite{Dal86}, which could be used to identify an individual with some \textit{sensitive attribute} (SA) such as his disease. 

\subsubsection{Data Privacy Attacks}

The most common privacy threats are \textit{record linkage}, \textit{attribute linkage}, and \textit{table linkage}, at which an attacker tries to link a record of an individual to a record in a published table, to a sensitive attribute in a published table, or to the published data table itself, respectively \cite{Fung10}. 

In record/attribute linkage, the attacker knows the victim’s record is in the released table. If some value on QI which matches victim’s QI, identifies a small number of records, the victim can be distinguished with high probability. In table linkage, however, the attacker does not know whether a victim exist in the released table and tries to determine the presence/absence of the victim’s record. In some other privacy models, we do not concern about such linkage attack, but we concern about the change in the attacker's probabilistic belief on the SA value of a victim after seeing the published data \cite{Fung10}.

\subsubsection{Data Privacy Models}

We explain some well-known approaches to prevent privacy attacks. The notion of \textit{$k$-anonymity} \cite{Samarati98} is a solution to record linkage attacks, where the QI of each record should be the same as at least $k$-1 other records. This ensures that the probability of linking an individual to a specific record based on QI is at most $\frac{1}{k}$. 

As a solution to attribute linkage attack, the \textit{$\ell$-diversity} notion \cite{Machanavajjhala06} requires each group of records with the same QI, to have at least $\ell$ ``well-represented'' SAs. This ensures that there are at least $\ell$ distinct values for the SA in each such group, thus automatically satisfies $k$-anonymity, where $k$=$\ell$. The $\ell$-diversity could not prevent attribute linkage attacks if the overall distribution of a SA is skewed. As a solution, the notion of \textit{$t$-closeness} \cite{Li07} requires the distribution of a sensitive attribute in any group on QID to be close to the distribution of the attribute in the overall table.

The $(\rho_{1}, \rho_{2})$-\textit{privacy} \cite{Evfimievski03} guarantees that if attacker's prior knowledge on a SA value  before data release is at most $\rho_1$ then after seeing the released data, his posterior knowledge is bounded by $\rho_2$, where 0$<$$\rho_1$$<$$\rho_2$$<$1. 

The notion of \textit{$\epsilon$-differential privacy} \cite{Dwork06} guarantees that the addition or removal of a ``single'' record in the database will not significantly change the statistical analysis results. $\epsilon$-differential privacy assures record owners that submitting their personal information to the database is very secure.

\subsubsection{Data Anonymization Techniques}

We explain three major techniques to guarantee privacy notions.

\begin{figure}
	\centering
	\includegraphics[scale=0.47]{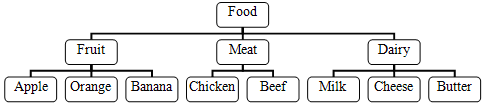}
    \caption{A simple taxonomy tree}
	\label{fig1}
\end{figure}

\textbf{Generalization and Suppression:} In suppression we delete some values, and in generalization we replace some values with their less specific values. For the generalization, we replace categorical attributes with respect to a given taxonomy, such as the one shown in Figure \ref{fig1}. Values in numerical attributes are usually replaced with an interval containing the original values. 

In \textit{full-domain generalization} \cite{Samarati01}, \cite{Sweeney02}, all values in an attribute are generalized to the \textit{same} level of the taxonomy tree. For example with taxonomy in Figure \ref{fig1}, if \textit{Beef} and \textit{Chicken} are generalized to \textit{Meat}, then \textit{Apple}, \textit{Orange} and \textit{Banana} should be generalized to \textit{Fruit}. In \textit{subtree generalization} \cite{Bayardo05}, \cite{Iyengar02}, either all child nodes or none are generalized. For example, in Figure \ref{fig1}, this scheme requires that if \textit{Beef} is generalized to \textit{Meat}, then the other child node, \textit{Chicken}, would also be generalized to \textit{Meat}, but \textit{Apple} and \textit{Orange}, which are child nodes of \textit{Fruit}, can remain ungeneralized. In \textit{cell generalization} \cite{Xu06}, also known as ``local recoding'', only some instances of a value will be generalized compared to ``global recoding'' in which if a value is generalized, all its instances are generalized. 

Major suppression techniques are \textit{Record suppression} \cite{Bayardo05}, \cite{Iyengar02}, \cite{Samarati01}, and \textit{cell suppression} (or local suppression)\cite{Cox80} \cite{Meyerson04}, which are processes of suppressing an entire record, or suppressing some instances of a given value in a database, respectively.

\textbf{Anatomization and Permutation:} In \textit{anatomization} \cite{Xiao06} the QI or the SAs are not modified and instead the QI data and the SAs data will be published in two separate tables: a QI table containing the quasi-identifier attributes, a SA table containing the sensitive attributes, and tables have one common \textit{GroupID} attribute.  In \textit{permutation} method \cite{Zhang07} records are partitioned into groups and then their SA values within each group will be shuffled.

\textbf{Perturbation and Randomization}: In \textit{perturbation} the original data values are replaced with some synthetic data values in such a way that the statistical information  is preserved.The \textit{additive noise} technique \cite{Adam89}\cite{Evfimievski02} alters a sensitive numerical data such as salary by adding a random value drawn from some distribution. The \textit{data swapping} method in which SA values of records are exchanged, can protect numerical and categorical attributes\cite{Reiss82}. Authors in \cite{Evfimievski03} also proposed a randomization approach based on data swapping to limit the attacker's background knowledge on inferring sensitive attributes.

\subsection{Contributions and Paper Organization}

In this survey, we provide an overview of the recent studies in privacy-preserving Web query log publishing. We explain privacy notions, attacks, and the utility challenges in query log anonymization. We categorize the recent privacy-preserving query log publishing techniques into transactional and non-transactional anonymity approaches. 

The rest of the paper is organized as follows. In Section 2, we study the problem of query log anonymization and its challenges. We categorize the existing anonymization methods in Section 3 and summarize and discuss these methods in section 4. We conclude the paper in Section 5.

\section{Privacy Preserving Web Query Log Data Publishing}

The problem of Web query-log anonymization have been examined with \cite{Kumar07}, \cite{Adar07}, and \cite{Cooper08} from Web community with focus on privacy attacks, and \cite{Evfimievski02}, \cite{Aggarwal07}, \cite{Ghi08}, \cite{Terrovitis08}, \cite{Xu08}, \cite{He09},  \cite{Korolova09}, \cite{MilaniFard10}, \cite{Jimmy10}, \cite{Cao10}, and \cite{Ero11} from the database community with focus on transaction database anonymization. In this survey we study both group of works and categorize them into non-transactional and transactional anonymity models respectively. However, the major part belongs to transactional model meaning that we treat query logs as \textit{transaction data} (unstructured data without a fixed set of attributes), where each transaction represents a query and each item represents a query term. Such data is a rich source for many data mining applications such as association rule mining, search recommendations, and etc.

\subsection{Privacy Attacks on Published Query Logs}

Authors in \cite{Chen09}, mentioned possible attacks on published Web queries. Some sensitive information can be obtained directly from query content such as social security numbers, credit card numbers, etc. Demographic and geographical information such as location, etc. could also help attacker to find the identity of a user. Even following the URL of pages that a user clicked can reveal the user’s identity when combined with his/her other queries. And finally, when a userID was identified, the adversary can easily discover all the user’s private queries by looking at the entire search history.

\subsection{Challenges in Anonymizing Query Logs}

As mentioned earlier, query log data can be seen as a special case of transaction data, where each transaction contains several ``items'' from an item universe. This item universe is typically very large, say thousands of items (such as catalog items in Amazon.com), and each transaction contains only a few items. If each item is treated as a binary attribute with 1/0 values, the transaction data is extremely high-dimensional and sparse. There are two main groups of challenges in anonymizing such high-dimensional data.

\subsubsection{Data Utility Challenge}

Query log (or transaction data) anonymization aims at preserving privacy while maintaining data utility and reducing the information loss. However, measuring the utility of the anonymized query logs is not always clear.
In case of suppression methods, the information loss can be a simple count of suppressed items. For generalization based techniques, various metrics have been proposed to measure the quality of generalized data including \textit{classification metric}, \textit{generalized loss metric} \cite{Iyengar02}, and \textit{discernibility metric} \cite{Bayardo05}. Some specific transaction anonymization loss measures are \textit{normalized centrality penalty} \cite{He09}, and \textit{group generalization distortion} \cite{MilaniFard10}. \textit{Itemset based utility} \cite{FungWang10} is another utility measure which captures frequent itemsets in transaction data.

Apart from the utility measures mentioned above, \cite{FungWang10} mentioned two other aspects for practical usefulness of the anonymized data. The first is the \textit{truthfulness} of results, i.e. the analysis results (such as support of frequent itemset) on the anonymized data holds on the original data. The second is the \textit{value exclusiveness}, i.e. the items in the modified data are exclusive of each other. This has a significant impact on many data mining tasks based on counting queries. For example, the local recoding transformation \cite{LeFevre06} does not have this property. Consider Figure 2, a local recoding can generalize some occurrences of ``Apple" and some occurrences of ``Orange", to ``Fruit". Now, it is not possible to count the number of transactions containing ``Apple" or ``Orange" from the modified data.

The major challenge for all query-log anonymization is reducing the significant information loss of the anonymized data.
This is because each dimension (any search term) could be potentially sensitive and a potential QID attribute used for record or attribute linkages, thus employing traditional privacy models, such as $k$-anonymity, would require including all dimensions into a single QID. Consequently lots of data has to be suppressed or generalized to the top-most values in order to satisfy $k$-anonymity, even for small values of $k$ \cite{Fung10}. Although removing sensitive terms based on the semantics of the search term and context can help increasing the utility of anonymized data, the removed sensitive terms can still be predicted based on user's other queries \cite{Jones07}.

\subsubsection{Data Privacy Challenge}

Anonymized query log data has some privacy issues which are even more important than the above utility issues. Firstly, the assumption of having an adversary with a very strong background knowledge can drastically affect the anonymized data utility. Therefore some researchers consider a \textit{bounded adversary} with a limited background knowledge (e.g. by a maximum number of items) \cite{FungWang10}. Although this assumption can be realistic, it does not hold for cases with unbounded adversary and thus privacy is breached.

Secondly, as discussed in \cite{Chen09}, an adversary can create multiple accounts and generate many queries using those accounts to create special query patterns (such as a lot of infrequent query, or a distinguishable signature), so that, when the search log is sanitized and released, the adversary can use those patterns to obtain private information about other users. Such issues are still not well studied.

\section{Web Query Log Anonymization Techniques}

We classify the query log anonymization methods into two groups which models query logs differently. The first group of works deals with query logs almost as is, while the second group treat query logs as a special case of high-dimensional transaction data. In this section we briefly explain previous representative works in each group. There might be some other very recent works which are to some extent an incremental variation of these methods.

\subsection{Non-Transactional Anonymity Models}

\subsubsection{Query Deletion and Hashing}

Seven query log privacy-enhancing techniques was discussed in \cite{Cooper08}, including deleting entire query logs, hashing query log content, deleting user identifiers, deleting personal information in query content, hashing user identifiers, shortening sessions, and deleting infrequent queries. 

Log deletion is the most privacy-enhancing technique; however, the utility of data drops to zero. Hashing queries is also not safe since other publicly available data, such as previously released query logs, or search engine statistics about queries in un-hashed form, can be used to pinpoint an individual. Similarly, hashing identifiers cannot guarantee eliminating the risk of privacy breach. 

Even after removing identifying information it may still be possible to link queries back to individuals by using other publicly available information. Although shortening sessions can be highly privacy-protective, due to removal of the link between a user and his/her entire query history, the query content may still contain identifying information, and thus the risks from accidental and malicious disclosure will not be totally resolved. In addition, query logs with short sessions are less useful for analysis. 

Deleting queries that appear infrequently in the logs was suggested in \cite{Adar07} as an effective way of removing queries that contain identifying information. Setting a threshold for being ``infrequent'' is however very challenging. Also studies showed that a large number of queries in huge query log datasets, occur a small number of times \cite{Bei04}. Consequently, this approach may lead to deletion of a remarkable amounts of non-identifying queries.

\subsubsection{Token based Hashing}

In \textit{token based hashing} \cite{Kumar07} a query is anonymized by tokenizing each query term and securely hashing each token to an identifier. One major problem with this technique is that if an unanonymized reference query log has been released previously, the adversary could apply co-occurrence analysis and frequency analysis on the reference query log to extract statistical properties of query terms and then processes the anonymized log to invert the hash function based on co-occurrences of tokens within queries. For example, if an adversary knows how often the query ``HIV treatment'' appears in a previously released log can use the statistics to decipher the separate hashes for ``HIV'' and ``treatment''.

\subsubsection{Secret Sharing and Split Personality}

The \textit{secret sharing} \cite{Adar07} anonymization method splits a query into $k$ random shares and publishes a new share for each distinct user issuing the same query. This technique guarantees $k$-anonymity because each share is useless on its own and all the $k$ shares are required to decode the secret. This means that a query can be decoded only when there are at least $k$ users issuing that query. The result is equivalent to suppressing all queries issued by less than $k$ users. Since queries are typically sparse, many queries will be suppressed as a result.

\textit{Split personality}, also proposed in \cite{Adar07}, focus on reducing the possibility of reconstructing search history of a user by splitting the logs of each user based on the user ``interests''.  For example, if a user is interested in both Sport and Art, then he will have two different profiles, one for the queries about Sport, and the other for the queries related to Art. In this way, the users become dissimilar to themselves, however the distortion makes it more difficult for researchers to correlate different facets. The authors, however, provided no formal privacy guarantee for this method.

\subsection{Transactional Anonymity Models}

The following works focused on transaction data anonymization. However, some also mentioned query log as a special case and used query log data in their experiments. 

\subsubsection{Randomization Methods}

An early work in transaction anonymizing was applying randomization methods, where some items are replaced with another and some "false" items are inserted into a transaction that looks like "true" items \cite{Evfimievski02}. Given a transaction $t$, the anonymized transaction $t$' is generated in three steps: The randomization operator selects $j$ items from $t$, uniformly at random with some probability (without replacement) and places them into $t$'. It considers each item not in $t$ and with some probability adds it to $t$'.

In another randomization approach \cite{Evfimievski03}, the authors proposed $(\rho_{1}, \rho_{2})$-\textit{privacy} which guarantees that if attacker's prior knowledge on a property $Q(t)$ of a transaction $t$ before data release is at most $\rho_1$ then after seeing the released randomized transaction $t$', his posterior knowledge is bounded by $\rho_2$, where 0$<$$\rho_1$$<$$\rho_2$$<$1. They presented a method for finding the perturbation probabilities that maximizes the expected value of $|t\cap t$'$|$  while ensuring $(\rho_{1}, \rho_{2})$-\textit{privacy}.

\subsubsection{Coherence Method}

The \textit{coherence} method \cite{Xu08} eliminates both record linkage attacks and attribute linkage attacks. The $(h, k, p)$-coherence privacy criterion requires that at least $k$ transactions must have any subset of at most $p$ non-sensitive items and at most $h$ percent of these transactions have some sensitive item. This ensures that, for an attacker with the power $p$, the probability of linking an individual to a transaction is limited to 1/$k$ and the probability of linking an individual to a sensitive item is limited to $h$.

Let $\beta$ denote the adversary's background knowledge that a transaction contains some non-sensitive items. An attack is modeled in the form of $\beta \rightarrow e$, where $e$ is a sensitive item. Let \textit{Sup}($\beta$) denote the support of $\beta$ i.e., the number of such transactions. $P(\beta \rightarrow e) = Sup(\beta\cup\{e\})/Sup(\beta)$ is the probability that a transaction contains $e$ given that it contains $\beta$. The \textit{breach probability} of $\beta$, denoted by $P_{breach}(\beta)$ is the maximum $P(\beta \rightarrow e)$ for any private item $e$. Assume an adversary's background knowledge is up to $p$ non-sensitive items, i.e., $|\beta|\leq p$. If \textit{Sup}$(\beta)<k$, the adversary is able to link an individual to a transaction (record linkage attack) and if $P_{breach}(\beta)> h$, the adversary is able to link an individual to a sensitive item (attribute linkage attack). 

A \textit{mole}, is any background knowledge (at most to the size $p$) that can result in a linking attack. \textit{Coherence} aim at eliminating all moles. For a setting of $(h, k, p)$, an itemset $\beta$, with $|\beta|$$\leq$$p$ and Sup($\beta$)$>$0, is called a mole if $Sup(\beta)<k$ or $P_{breach}(\beta)>h$. The data $D$ is $(h, k, p)$-coherent if $D$ contains no moles. Authors in \cite{Xu08} applied the \textit{total item suppression} technique to enforce $(h, k, p)$-coherence. Total suppression of an item refers to deleting the item from \textit{all} transactions containing it. Although total suppression results in a high information loss when the data is sparse, it has two nice properties: (1) eliminating all moles containing the suppressed item, and (2) keeping the support of any remaining itemset, equal to the support in the original data. The latter one implies that any result derived from the modified data, also holds on the original one which is not hold for \textit{partial suppression}.

Since an optimal solution to $(h,k,p)$-coherence, i.e. with minimum information loss (suppressed items) is \textit{NP}-hard \cite{Xu08}, authors proposed a heuristic solution. They defined \textit{minimal moles} as those moles that contain no proper subset as a mole in which removing them is sufficient for removing all moles. An algorithm similar to the well-known \textit{Apriori} algorithm for mining frequent itemsets, was used to find all minimal moles.

One problem with the coherence method is the scalability issue considering exponential growth of itemsets. In a similar work, \cite{Xu08a} suggested using sets of maximal and minimal itemsets, called borders. These borders are typically much smaller than the full sets of all itemsets that they represent, thus their solution requires much less space and time.

\subsubsection{Band Matrix Approach}

An anonymization method is proposed in \cite{Ghi08} to prevent attribute linkage attacks for high-dimensional data with sensitive items, using a \textit{band matrix} technique. In a band matrix, non-zero entries are confined to a diagonal band and zero entries on either side. In such a matrix, rows correspond to transactions and columns correspond to items, with the 0/1 value in each entry. In their method, items are divided into \textit{sensitive items}, and \textit{non-sensitive items}. A non-sensitive transaction, is a transaction with no sensitive items and sensitive transactions are those with at least one sensitive item. 

A transaction set $T$ has \textit{privacy degree} of $p$ if the probability of associating any transaction $t$$\in$$T$ with a particular sensitive item does not exceed $\frac{1}{p}$. To achieve this privacy requirement, \cite{Ghi08} suggested applying two phases: (1) transforming the data to a band matrix (using \textit{Reverse Cuthill-McKee} algorithm) with respect to non-sensitive attributes, and (2) grouping each sensitive transaction with non-sensitive transactions or sensitive ones with different sensitive items. The intuition why such band matrix formation is helpful, is that it organizes data such that consecutive transactions are very likely to share many common non-sensitive items and this results in a smaller reconstruction error.

In the second phase each sensitive transaction will be grouped with non-sensitive transactions or sensitive ones with different sensitive items. A greedy algorithm based on the ``one-occurrence-per-group'' heuristic, was proposed in \cite{Ghi08} which allows only one occurrence of each sensitive item in a group.

\subsubsection{$k^m$-Anonymity}

To address the record linkage attacks in transaction data, \cite{Terrovitis08} proposed the $k^m$-anonymity notion which assumes that any subset of items can be used as background knowledge. In this method unlike \textit{coherence} and \textit{band matrix} approach, data is not distinguished as sensitive and non-sensitive, but it is considered both as potential quasi-identifiers and potential sensitive data. It assumes, like the \textit{coherence} method, that an adversary knows at most $m$ number of items as background knowledge.

A transaction database is $k^m$-anonymous if for any set of up to $m$ items, there exist at least $k$ transactions that contain those items in the published database. We can consider this privacy notion as a special case of $(h, k, p)$-coherence with $h$ = 100\% and $p$ = $m$, meaning that a subset of items that causes violation of $k^m$-anonymity is a mole under the
$coherence$ model.

The anonymization method in \cite{Terrovitis08} applies generalization in form of the \textit{global recoding} scheme in which when a child node is generalized to its parent, all its sibling nodes will also be generalized to their parent node, and the generalization process is applies to all transactions in the database. Each generalization corresponds to a possible horizontal \textit{cut} of the taxonomy tree. The information loss of a cut is measured using the \textit{normalized certainty penalty} loss metric \cite{Xu06} which captures the degree of generalization of an item $i$, by considering the percentage of leaf nodes under $i$ in the item taxonomy.

If a cut results in a $k^m$-anonymous database, then all its more general cuts, also result in a $k^m$-anonymous database. This is called the \textit{monotonicity property} of cuts \cite{Terrovitis08}. The $k^m$-anonymization problem is to find a $k^m$-anonymous transformation with the minimum information loss. Based on the monotonicity property and in order to prevent higher information loss, as soon as we find a cut that satisfies the $k^m$-anonymity constraint, we do not have to find a more general cut.

Generating the set of all possible cuts and checking the anonymity violation for every subset of up to $m$ items is not applicable for large, realistic problems. Thus, authors proposed a greedy heuristic method called \textit{Apriori anonymization} (\textit{AA}) which is based on the apriori principle: if an itemset $J$ of size $i$, violates the anonymity requirement, then each superset of $J$ also violates the anonymity requirement. It explores the space of itemsets in an apriori, bottom-up scheme. Meaning that before checking if $\ell$-itemsets ($\ell$ = 2,$\dots$, $m$) violates the anonymity requirement, we first eliminate the possible anonymity violation caused by ($\ell$-1)-itemsets. This method drastically reduces the number of itemsets that must be checked at a higher level, since detailed items could have been generalized.

\subsubsection{Transactional $k$-Anonymity}

The assumption of bounded background knowledge of an adversary in the \textit{coherence} and the \textit{$k^m$-anonymity} methods, has two limitations. Firstly, in many cases it is not possible to determine this bound in advance. Secondly, these methods can ensure $k$-anonymity (with $p$ or $m$ set to the maximum transaction length in the database) only if adversary's background knowledge is limited to the \textit{presence} of items. If the background knowledge is on the ``absence'' of items, the adversary may exclude transactions and focus on fewer than $k$ transactions. For example, consider an adversary who knows that Bob has bought ``Orange" and ``Chicken", but has \textit{not} bought ``Milk". Suppose that three transactions contain ``Orange", and ``Chicken", in which two of them contain ``Milk". The adversary can exclude the two transaction containing ``Milk" and link the remaining transaction to Bob. Here, $k^m$ privacy with $k$=2 and $m$=3 is violated, even by setting $m$ to the maximum transaction length. 

The $k$-anonymity approach in \cite{He09}, which we refer to as the $Partition$ method, avoids this problem since all transactions in the same equivalence class are identical. They extended the original $k$-anonymity for relational data \cite{Samarati01} \cite{Sweeney02}, to the transactional $k$-anonymity for ``set-valued data'', in which a set of values are associated with an individual. A transaction database $D$ is $k$-anonymous if every transaction in $D$ occurs at least $k$ times. Authors in \cite{He09} showed that every database which satisfies $k$-anonymity, also satisfies $k^m$-anonymity for all $m$ values, however, the reverse does not always hold. 

The $Partition$ method is the extended version of the top-down \textit{Mondrian} \cite{LeFevre06} algorithm for relational data. In this method, if several items are generalized to the same item, only one occurrence of the generalized item will be kept in the generalized transaction. It starts with the single partition containing all transactions with all items generalized to the \textit{root} item. Then it recursively splits a partition by specializing a node in the taxonomy for all the transactions in the partition. Next all the transactions in the partition with the same specialized item are distributed to the same sub-partition. At the end of distribution, some small sub-partitions with less than $k$ transactions are merged into a special leftover sub-partition to be redistributed. The partitioning stops if $k$-anonymity condition is violated. Unlike the \textit{Apriori} anonymization \cite{Terrovitis08}, the \textit{Partition} approach follows a \textit{local recoding} scheme.

\subsubsection{Clustering-Based $k$-Anonymity}

The $Partition$ method suffers from significant information loss for two reasons. Firstly, it stops partitioning the data at a high level of the taxonomy because the exponential branching for generating sub-partitions quickly diminishes the size of a sub-partition and causes $k$-anonymity violation. This is especially true for query logs with large and diverse item universe. Secondly, it does not deal with item duplication in the generalized transaction. In fact preserving term frequency (as much as possible) is an important issue for many applications such as TFIDF used by ranking algorithms.

Authors in \cite{MilaniFard10}, adopted the the privacy notion of transactional $k$-anonymity \cite{He09} and proposed a clustering approach to query log anonymization as a solution to the above shortcomings of the $Partition$ method. The main idea in \cite{MilaniFard10} is grouping ``similar'' transactions together, to reduce the amount of required generalization and suppression to make them identical. For example, the generalized transaction for $<$\textit{Apple}$>$ and $<$\textit{Milk}$>$ is $<$\textit{Food}$>$, and for $<$\textit{Apple}$>$ and $<$\textit{Orange}$>$ is $<$\textit{Fruit}$>$. Clearly the former entails more information loss. Therefore, the transaction anonymization can be treated as a clustering problem such that each cluster must contain at least $k$ transactions and these transactions should be ``similar''. 

They defined a transaction as a bag of items (thus allowing duplicate items). A transaction $t$' is a \textit{generalized transaction} of a transaction $t$, if each item $i$'$\in$$t$' represents (the generalization of) one ``distinct'' item $i \in t$. This transaction model has two distinctions from \cite{He09}. First, it allows duplicate items in a transaction and in its generalized transaction. For example if $t$'=$<$$Fruit$, $Fruit$$>$ is a generalized transaction of $t$, $t$' represents two leaf items under $Fruit$ in $t$. Second, it allows items in a transaction to be on the same path in the item taxonomy while each item represents a distinct leaf item. For example, we interpret the transaction $<$$Fruit$, $Food$$>$ as: $Fruit$ represents a leaf item under $Fruit$ and $Food$ represents a leaf item under $Food$ that is not represented by $Fruit$. 

The \textit{least common generalization} $(LCG)$ was proposed as a way to measure the similarity of a subset of transactions. The $LCG$ of a set of transactions $S$, is a common generalized transaction for all of the transactions in $S$, and there is no other more special common generalized transaction. The authors devised an efficient linear-time bottom-up item generalization algorithm to compute $LCG$. Authors proposed \textit{group generalization distortion} $(GGD)$ as a measure to capture both generalization and suppression distortion of a set of transactions. They formulated the transaction anonymization as the problem of clustering a set of transactions into clusters of size at least $k$ such that the sum of $GGD$ of \textit{LCG} of all clusters is minimized. Since the problem is $NP$-hard, they presented a heuristic linear-time algorithm, called $Clump$, which unlike $Partition$, preserves duplicate items after generalization. 

\subsubsection{Heuristic Generalization with Heuristic Suppression}

Authors  in \cite{Jimmy10} were motivated by the limitations of the $k^m$-anonymity, and proposed to integrate the global generalization technique in \cite{Terrovitis08} with the total item suppression technique in \cite{Xu08} for enforcing $k^m$-anonymity. They applied \textit{full subtree generalization} technique \cite{Terrovitis08}, meaning that a generalization solution $Cut$ is defined by a cut on a taxonomy tree with exactly one item on every root-to-leaf path. Since the full subtree generalization can suffer from excessive distortion in the presence of outliers, suppressing a few outlier items will reduce information loss caused by high amount of generalization. They applied \textit{total item suppression} technique, which removes some items of $Cut$ from all transactions. The loss metric is the aggregate of both generalization and suppression.

The anonymized data is derived in two steps: first the items are generalized with respect to the $Cut$ and then some items of the $Cut$ are suppressed in all transactions. Since the number of cuts for a taxonomy is exponential in the number of items, enumerating suppression scenarios for a cut is also intractable. Consequently, authors provided a heuristic approach to address this issues.

\subsubsection{Sketch-based Anonymization}

The sketch-based privacy-preserving approach \cite{Aggarwal07} reduces the dimensionality of the data by producing a much smaller number of features to represent the data. This technique is specifically effective for high-dimensional sparse data such as query logs. The idea is to replace a user's search history by a set of sketches. Two privacy criteria associated with this technique are $\delta$-anonymity and $k$-variance.

The $\delta$-\textit{Anonymity} ensures that the uncertainty in the reconstructed value of each term frequency is at least $\delta$. As noted in \cite{Aggarwal07}, a disadvantage of $\delta$-anonymity is that it treats each user independently regardless of whether there are other users similar to him/her. They argued that it is desirable to give outliers (users who use unique terms) more protection than users who are similar to many others. Thus, they define the $k$-variance which ensures that any user's sanitized search history cannot be easily distinguished from its $k$-nearest neighbors. They described algorithms for $\delta$-anonymity and $k$-variance using suppression.

\subsubsection{Semantic Microaggregation}

The anonymization method in \cite{Ero11}, clusters the queries and then replace the original queries by the centroids of the corresponding clusters considering the \textit{semantics} of the queries. Authors argued that creating a cluster with queries from users with different ``interests'' can result in useless protected logs and thus queries of users with common interests between them should be grouped in the same cluster. 

They used \textit{Open Directory Project} \footnote{http://www.dmoz.org/} to compute the semantic distances between users and partition queries into groups of $k$ users with similar interests. In the aggregation phase, they compute a new user as the representative (or centroid) of the cluster, which summarizes the queries of all the users of the cluster. The query items for the centroid are selected by a probabilistic approach based on the contribution of the user with respect to number of transactions in the cluster.

\subsubsection{Differential Privacy}

\textit{Differential privacy} \cite{Dwork06} is one of the state-of-the-art techniques for ensuring privacy and is more robust to attacks than any other existing privacy definitions. The notion of differential privacy was applied for search queries in \cite{Korolova09} which adds a random noise to any statistic of a search log such as a term frequency. This random noise is drawn independently from the \textit{Laplace distribution} with mean zero and a scaling parameter. The algorithm output contains frequent queries with noisy statistic of the queries and the clicked URLs.

\subsubsection{$\rho$-Uncertainty}

The privacy notion \textit{$\rho$-uncertainty} \cite{Cao10} ensures that the confidence of any sensitive association rule is at most $\rho$, while truthful association rules can still be derived. Like the works in \cite{Ghi08}, and \cite{Xu08}, they distinguish between public (non-sensitive) and private (sensitive) items. Formally, a $\rho$-uncertain transaction set $D$ does not allow an attacker knowing any subset of a transaction $t$$\in$$D$ to infer a sensitive item in $t$ with confidence higher than $\rho$. 

The authors proposed a technique that combines global generalization over non-sensitive items and selective global suppression of some items. This notion is similar to $(h,k,p)$-coherence, however, the $\rho$-uncertainty model allows an adversary with some prior knowledge on the private items.

\section{Summary and Discussion}

\subsection{Privacy Preservation}
In this survey we considered two models for query log anonymization: non-transactional model and transactional model. Although the techniques mentioned in non-transactional model in Section 3.1 protect privacy to some extent, there is a lack of formal privacy guarantee. For example, the release of the AOL query log still leads to the re-identification of a search engine user even after hashing user’s identifiers \cite{Barbaro06}. This is because the query content itself may be used together with publicly available information for linking attacks.

In the transactional model we consider Web query logs as unstructured transaction data and therefore focus on query-log anonymization from transaction database anonymization point of view. Such a modeling, however, might not be a good idea since there are strong correlations between keywords within a query (based on natural language), and between queries within a single session. This is not true of transactions. Utilizing these correlations can help develop better solutions for the problem in the future.

Among the previous work in transaction anonymization, the \textit{coherence} \cite{Xu08} approach can both prevent record linkage attacks and attribute linkage attacks. \textit{band matrix} \cite{Ghi08} and \textit{$\rho$-uncertainty}\cite{Cao10} approach can prevent attribute linkage attack, and both $k^m$-anonymization \cite{Terrovitis08}  \cite{Jimmy10} and $k$-anonymization \cite{He09} \cite{MilaniFard10} prevent record linkage attack. 

Both $k^m$-anonymization and $k$-anonymization, do not distinguish data as sensitive and non-sensitive but as potential QI and SA. In fact, determining which items are sensitive is not always possible in many real applications considering huge size of the item universe. The adversary's background knowledge is bounded in \textit{coherence} and $k^m$-anonymization, while in \textit{band matrix} and $k$-anonymization we do not limit the attacker's knowledge. A security issue about bounded knowledge in \textit{coherence} and $k^m$-anonymization was explained by \cite{He09} that if background knowledge is on the ``absence'' of items, the attacker may exclude transactions using this knowledge and focus on fewer than $k$ transactions. The $HgHs$ approach also has this privacy issue.

For the \textit{sketch-based} privacy-preserving approach \cite{Aggarwal07}, authors in \cite{Chen09} argued that one should be careful of releasing the (pseudo)randomly generated values that were used in the sanitization process in \cite{Aggarwal07} since this may allow attackers to reconstruct the original data which is a privacy breach.

While applying differential privacy for search queries \cite{Korolova09} is very promising, like every existing privacy definition, it is susceptible to active attacks. The assumption that users behave honestly may lead to privacy breach. If an attacker creates multiple accounts and in some of his first queries issues a private query such as someone else's credit card number, it could result in publishing this private data by the search engine.

\subsection{Utility Preservation}

As discussed in Section 2.2.1, important utility factors for the anonymized data are \textit{item generalization/suppression loss}, \textit{truthfulness}, \textit{itemset utility}, \textit{value exclusiveness}, and \textit{item frequency}. 

Authors in \cite{Xu08a} and \cite{Xu08} assume that the taxonomy tree for transaction data tend to be flat and fanout, and thus decided to use item suppression instead of generalization. In this case, employing generalization loses more information than employing item suppression. However, if the transaction database is too sparse, then the item suppression of the coherence may cause a large information loss. If the data is sparse and the taxonomy is ``slim" and ``tall", the generalization scheme in $k^m$-anonymization and the $k$-anonymization  could work better, while if the taxonomy is ``short" and ``wide", generalization causes larger information loss \cite{FungWang10}\cite{MilaniFard10}. 

Data analysis on anonymized data is considered \textit{truthful} with respect to the original data if the analysis results obtained from the modified data holds on the original data \cite{FungWang10}. The \textit{coherence} approach and $k^m$-anonymity approaches guarantee truthful analysis while it is not the case for the $k$-anonymization.

The analysis of frequent itemsets \cite{Agrawal93}, i.e., the items that co-occur frequently in transactions, has a vast application in data mining applications such as association rule mining, search recommendations, and etc. Thus preserving itemsets is an important utility factor. Among the discussed approaches, coherence and band matrix can preserve such itemset utility.

The local generalization in transactional $k$-anonymity approach has a smaller information loss than global generalization, however, the anonymized data does not have the \textit{value exclusiveness}, which is important to preserve for many data mining algorithms. This means that new algorithms must be designed to analyze such data \cite{FungWang10}.

Most of the previous works in transaction data anonymization do not deal with item duplication meaning that the frequency of a term in a query can not be preserved well and will affect utilities such as count query results. For example the information loss of $k$-anonymity method in \cite{He09} can be high due to item generalization, and eliminating duplicate generalized item. The latter reason of information loss was not measured by an usual information loss metric for relational data where no attribute value will be eliminated by generalization. Authors in \cite{MilaniFard10}, however, designed their anonymization method in such a way which preserves item frequency.

There is no guarantee for minimum data distortion in the semantic microaggregation technique \cite{Ero11} while computing the centroid for the clusters. Moreover, authors did not consider item generalization and its cost in their model. For The \textit{sketch-based} privacy-preserving approach \cite{Aggarwal07}, it is interesting to see if it would be useful for anonymizing real search logs, and when we only have sanitized search logs, what kinds of search log analysis can still be conducted with acceptable accuracy.

\section{Conclusion}

Publishing Web query logs for research/marketing is restricted by privacy concerns. On the other hand, achieving a suitable trade off between privacy and utility of query log data is a challenging problem. We surveyed some recent studies on query log anonymization and categorized them into two groups based on how they treat queries. Major works consider query logs as transaction data and apply techniques to guarantee a desired level of privacy. While there is progress in privacy-preservation of published query logs, preserving data utility is still a challenging issue.



\begin{thebibliography}{99}
\small{

\bibitem{Adam89} N. R. Adam, and J. C. Wortman. Security control methods for statistical databases, In ACM Computer Surveys, V 21, 4 (December), pages 515--556, 1989.

\bibitem{Adar07} E. Adar. User 4XXXXX9: Anonymizing query logs. In WWW Query Log Workshop, 2007.

\bibitem{Aggarwal07} C. C. Aggarwal and P. S. Yu. On privacy-preservation of text and sparse binary data with sketches. In SDM, pages 57--67, 2007.

\bibitem{Agrawal00} R. Agrawal, and R. Srikant. Privacy preserving data mining, In ACM SIGMOD, pages 439--450, 2000.

\bibitem{Agrawal93} R. Agrawal, T. Imielinski, and A. N. Swami. Mining association rules
between sets of items in large databases. In SIGMOD, pages 207--216, 1993.

\bibitem{Barbaro06} M. Barbaro, and T. Zeller. A face is exposed for AOL searcher no. 4417749. The New York Times, 2006-08-09

\bibitem{Bayardo05} R. J. Bayardo, and R. Agrawal. Data privacy through optimal k-anonymization. In ICDE, pages 217--228, 2005.

\bibitem{Bei04} S. Beitzel, E. Jensen, A. Chowdhury, D. Grossman, and O. Frieder. Hourly analysis of a very large topically categorized Web query log, In ACM SIGIR, pages 321--328, 2004.

\bibitem{Cao10} J. Cao, P. Karras, C. Raissi, and K. Tan. $\rho$-uncertainty: Inference-proof transaction anonymization. In VLDB, pages 1033--1044 , 2010.

\bibitem{Cooper08} A. Cooper. A survey of query log privacy enhancing techniques from a policy perspective. In ACM Transactions on the Web, V. 2, No. 4, ACM, pages 1--27, 2008.

\bibitem{Cox80} L. H.  Cox. Suppression methodology and statistical disclosure control, In Journal of the American Statistical Association, V 75, 370, pages 377--385, 1980.

\bibitem{Chen09} B. Chen, D. Kifer, K. LeFevre, and A. Machanavajjhala. Privacy-Preserving Data Publishing. In Foundations and Trends in Databases, Volume 2 Issue 1–2, pages 1--167, January 2009.

\bibitem{Dal86} T. Dalenius. Finding a needle in a haystack - or identifying anonymous census record. In Journal of Official Statistics, V2, pages 329--336, 1986.

\bibitem{Du03} W. Du, and Z. Zhang. Using randomized response techniques for privacy-preserving data mining, In ACM SIGKDD, pages 505--510, 2003.

\bibitem{Dwork06} C. Dwork. Differential privacy. In the Intl. Colloquium on Automata, Languages and Programming (ICALP), pages 1--12, 2006.

\bibitem{Ero11} A. Erola, J. Castella-Roca, G.  Navarro-Arribas, and V. Torra. Semantic microaggregation for the anonymization of query logs. In Privacy in Statistical Databases, volume 6344 of Lecture Notes in Computer Science, pages 127--137, 2011.

\bibitem{Evfimievski02} A. Evfimievski, R. Srikant, R. Agrawal, and J. Gehrke. Privacy preserving mining of association rules. In SIGKDD, pages 217--228, 2002.

\bibitem{Evfimievski03} A. Evfimievski, J. Gehrke, and R. Srikant. Limiting privacy breaches in privacy preserving data mining, In ACM PODS, 211--222 , 2003.

\bibitem{Fung10} B. C. M. Fung, K. Wang, R. Chen, P.S. Yu.  Privacy preserving data publishing: a survey on recent developments. ACM Computing Surveys, V. 42, Issue No 4, pages 14:1--14:53, 2010.

\bibitem{FungWang10} B. C. M. Fung, K. Wang, A. W.-C. Fu, and P. S. Yu. Introduction to Privacy-Preserving Data Publishing: Concepts and Techniques, Data Mining and Knowledge Discovery Series. Chapman \& Hall/CRC, ISBN: 9781420091489, 2010.

\bibitem{Ghi08} G.  Ghinita, Y.  Tao, and  P. Kalnis. On the anonymization of sparse high-dimensional data, In ICDE, pages 715--724 , 2008.

\bibitem{Hafner06} K. Hafner, T. Zeller. Tempting data, privacy concerns; researchers yearn to use AOL logs, but they hesitate. The New York Times. 2006-09-13

\bibitem{He09} Y. He, J. F. Naughton. Anonymization of set valued data via top-down, local generalization. In VLDB, pages  934--945, 2009.
 
\bibitem{Iyengar02} V. S. Iyengar. Transforming data to satisfy privacy constraints. In SIGKDD, pages 279--288 , 2002.

\bibitem{Jones07} R. Jones, R. Kumar, B. Pang, and A. Tomkins. I know what you did last summer -- query logs and user privacy. In CIKM, pages 909--914, 2007.

\bibitem{Korolova09} A. Korolova, K. Kenthapadi, N. Mishra, and A. Ntoulas. Releasing search queries and clicks privately. In WWW, pages 171--171, 2009.

\bibitem{Kumar07} R. Kumar, J. Novak, B. Pang, and A.Tomkins. On anonymizing query logs via token-based hashing. In WWW, pages 629--638, 2007.

\bibitem{LeFevre06} K. LeFevre, D. J. DeWitt, and R. Ramakrishnan. Mondrian multidimensional k-anonymity, In ICDE, pages 25, 2006.

\bibitem{Li07} N. Li, T. Li, and S. Venkitasubramaniam. $t$-closeness: Privacy beyond $k$-anonymity and $\ell$-diversity, In IEEE ICDE, pages 106--115 , 2007.

\bibitem{Jimmy10} J. Liu,, and K. Wang. Anonymizing transaction data by integrating suppression and generalization, In PAKDD, pages 171--180, 2010.

\bibitem{Machanavajjhala06} A. Machanavajjhala, J. Gehrke, D. Kifer, and M. Venkitasubramaniam. $\ell$-diversity: Privacy beyond $k$-anonymity, In IEEE ICDE, 2006.

\bibitem{Meyerson04} A. Meyerson, and R. Williams. On the complexity of optimal k-anonymity. In PODS, pages 223--228, 2004.

\bibitem{MilaniFard10} A. Milani Fard, and K. Wang. An effective clustering approach to web query log anonymization. In Intl. Conf. on Security and Cryptography (SECRYPT), pages 109--119, 2010.

\bibitem{MilaniFard10Thesis} A. Milani Fard. Clustering-based Web Query Log Anonymization. M.Sc. Thesis, Simon Fraser University, 2010.

\bibitem{Reiss82} S. P. Reiss, M. J. Post, and T. Dalenius. Non-reversible privacy transformations, In ACM PODS, pages 139--146, 1982.

\bibitem{Samarati98} P. Samarati and L. Sweeney. Generalizing data to provide anonymity when disclosing information, In ACM SIGACT-SIGMOD-SIGART PODS, page 188, 1998.

\bibitem{Samarati01} P. Samarati. Protecting respondents' identities in microdata releases. IEEE Transactions on Knowledge and Data Engineering (TKDE), V. 13, pages 1010--1027, 2001.

\bibitem{Sweeney02} L. Sweeney. Achieving k-anonymity privacy protection using generalization and suppression. International Journal of Uncertainty, Fuzziness and Knowledge-Based Systems, Volume 10 Issue 5, pages 571--588, October 2002.

\bibitem{Terrovitis08} M. Terrovitis, N. Mamoulis, and P. Kalnis. Privacy preserving anonymization of set valued data, In VLDB,pages 115--125, 2008.

\bibitem{Xiao06} X. Xiao, and Y. Tao. Anatomy: Simple and effective privacy preservation, In VLDB, pages 139--150 , 2006.

\bibitem{Xu06} J. Xu, and W. Wang, J. Pei, X. Wang, B. Shi, and A. W. Fu. Utility-based anonymization using local recoding, In ACM SIGKDD, pages 785--790, 2006.

\bibitem{Xu08a} Y. Xu, B. Fung, K. Wang, A. Fu, and J. Pei. Publishing sensitive transactions for itemset utility. In ICDM, pages 1109--1114, 2008.

\bibitem{Xu08} Y. Xu, K. Wang, A. Fu, and P. S. Yu. Anonymizing transaction databases for publication, In ACM SIGKDD, pages 767--775, 2008.

\bibitem{Zhang07} Q. Zhang, N. Koudas, D. Srivastava, and T. Yu. Aggregate query answering on anonymized tables, In IEEE ICDE, pages 116--125 , 2007.

}
\end{thebibliography}


\end{document}